\documentclass[10pt,journal,compsoc]{IEEEtran}                 

\ifCLASSOPTIONcompsoc
\usepackage[nocompress]{cite}
\else
\usepackage{cite}
\fi

\usepackage{microtype}                 
\PassOptionsToPackage{warn}{textcomp}  
\usepackage{textcomp}                  
\usepackage{mathptmx}                  
\usepackage{times}                     
\usepackage{cite}                      
\usepackage{tabu}                      
\usepackage{booktabs}                  

\usepackage{todonotes}
\usepackage{tabu}
\usepackage{enumitem}
\usepackage{caption}
\usepackage{subcaption}
\usepackage{cite}
\usepackage{amsmath}
\usepackage{xspace}

\usepackage{xcolor}
\usepackage[draft,inline,nomargin,index]{fixme}

\newcommand{\exhdtored}{\ref{ex:hd2d}-2DReduction\xspace}
\newcommand{\exhypp}{\ref{ex:hypp}-Parameters\xspace}
\newcommand{\exvecsz}{\ref{ex:vecsz}-Dimensions\xspace}
\newcommand{\exoverfit}{\ref{ex:overfit}-Overfit\xspace}
\newcommand{\excorpora}{\ref{ex:corpora}-Corpora\xspace}
\newcommand{\exlanguages}{\ref{ex:languages}-Languages\xspace}

\newcommand{\sysname}{\emph{embComp}\xspace}

\newcommand\crule[3][black]{\textcolor{#1}{\rule{#2}{#3}}}

\definecolor{tableleft}{rgb}{0.278, 0.427, 0.616}
\definecolor{tableright}{rgb}{0.675, 0.447, 0.522}

\definecolor{countscolor}{rgb}{0.133, 0.455, 0.639}
\definecolor{rankscolor}{rgb}{0.541, 0.251, 0.580}
\definecolor{distancescolor}{rgb}{0.204, 0.514, 0.290}

\begin{document}

\title{\sysname: Visual Interactive Comparison of Vector Embeddings}

\author{Florian Heimerl,  Christoph Kralj,  Torsten M\"oller and Michael Gleicher}

\IEEEtitleabstractindextext{

	\begin{center}
		\vspace{-3mm}
		\hspace{-6mm}
		\captionsetup{type=figure}\setcounter{figure}{0}
		\includegraphics[width=.95\textwidth]{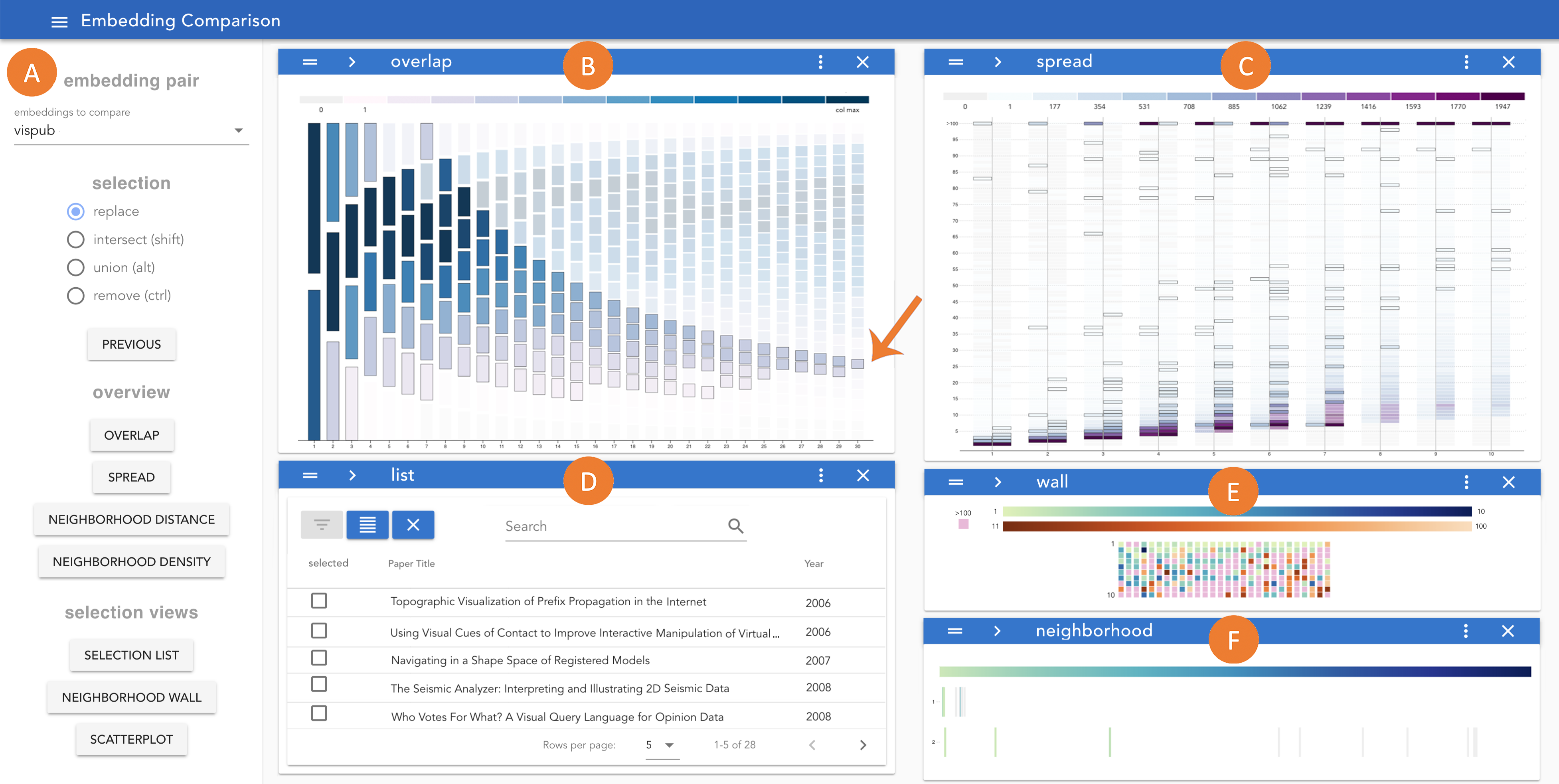}
		\captionof{figure}{ The \sysname system on the example of Section 3.3.
			The (A) \emph{configuration drawer} allows users to configure the desktop, select datasets and manage selections.
			On the right, the desktop contains five views:
			(B) \emph{neighborhood overlap view},
			(C) \emph{neighborhood spread view},
			(D) \emph{selection list view},
			(E) \emph{neighborhood wall view},
			and
			(F) \emph{neighborhood sequence view}.
			In this screenshot, the user has selected a bin in the \emph{overlap view} on the bottom right (marked by the arrow), adding the objects it contains to the selection set.
			Other views highlight bins that contain selected objects (non-transparent bins with dark outlines in B and C), or show a representation of each selected object (D and E).
			}\label{fig:teaser}
	\end{center}

	\begin{abstract}
		This paper introduces \sysname, a novel approach for comparing two embeddings that capture the similarity between objects, such as word and document embeddings.
We survey scenarios where comparing these embedding spaces is useful.
From those scenarios, we derive common tasks, introduce visual analysis methods that support these tasks, and combine them into a comprehensive system.
One of \sysname's central features are overview visualizations that are based on metrics for measuring differences in the local structure around objects.
Summarizing these local metrics over the embeddings provides global overviews of similarities and differences.
Detail views allow comparison of the local structure around selected objects and relating this local information to the global views.
Integrating and connecting all of these components, \sysname supports a range of analysis workflows that help understand similarities and differences between embedding spaces.
We assess our approach by applying it in several use cases, including understanding corpora differences via word vector embeddings, and understanding algorithmic differences in generating embeddings.
	\end{abstract}
}

\maketitle

\IEEEdisplaynontitleabstractindextext

\IEEEpeerreviewmaketitle

\IEEEraisesectionheading{\section{Introduction}\label{sec:intro}}

Analysis techniques for complex objects, such as documents, images, or words, often rely on \emph{object embeddings} that associate each object with a vector in a latent space.
These embeddings are constructed by placing the objects into a vector space such that related objects are close.
Embeddings may be examined directly, or used as input to other analytic steps that use distance information, such as clustering and nearest-neighbor classification.

Different embedding methods and parameterizations of these methods can lead to significantly different results, and embeddings can encode a variety of relationships in different ways.
Comparison between embeddings is thus an essential part of the embedding workflow.
It can be useful for model selection, model tuning, and understanding the underlying data set.
For all those goals, users need to understand the differences between embeddings, and investigate potential causes and effects of these differences.
This makes embedding comparison a complex, user-centric problem that is not well understood so far.

Embeddings position a potentially large number of objects within a given space.
The results may differ significantly with respect to the absolute position of the objects, but relative positions and local neighborhoods are typically most relevant.
With this work, we enable users to see overviews of local correspondences between embeddings and enable them to drill down to details to inspect differences on a more fine-grained level.
In contrast, existing approaches to comparing embeddings typically summarize similarities and differences with one single value, such as stress~\cite{kruskal1964multidimensional}, or examine a set of pre-selected examples~\cite{Heimerl2018}.
The former is problematic because it is a gross oversimplification of the measured difference, while the latter requires the identification of examples and fails to provide a broader context.

Comparing embeddings is challenging because it must consider similarities and differences in the \emph{relationships} between objects in different embeddings.
For example, two objects may be close together even if the embeddings have a different number of dimensions or were created by different algorithms.
In other words, both objects could have very different positions in these embeddings, but still be close in distance.
Embedding comparison tools must help the viewer understand the similarities and differences in the relationships between objects, independent of where those objects are placed in their respective spaces.
These challenges are exacerbated because embeddings often involve large collections of high dimensional objects.

In this paper, we provide a novel, visual approach to comparing object embeddings.
Our approach uses overviews that provide a synopsis of differences in local object relationships over the data set, selection mechanisms that use the summaries to identify interesting subsets of objects, and local views that help understand individual relationships and relate them back to the global views.
A variety of local metrics capture different aspects of object relationships.
Overview visualizations show the distribution of these metrics over the collection of objects in the embeddings.
These overviews allow the user to assess the overall amount of difference, and to identify sets of objects with potentially interesting behavior.
Identified subsets of objects can be examined more closely to show commonalities and differences in the selection detail views, their relationships with nearby objects in the local views, and their context within the larger embedding.

Our contributions are: (1) a visual interactive approach to embedding comparison and an implementation thereof; (2) a collection of metrics that convey differences between embeddings and corresponding visual designs for distributions of those metrics for a pair of embeddings; (3) the specific designs for some of its components; and (4) a flexible selection mechanism based on set algebra.
The presented approach is general: it can be applied to any type of object embedding where pairwise distances between objects are important.
We evaluate our approach by showing how it enables a variety of uses, and demonstrate the benefits of systematic comparison of embeddings.

\begin{figure*}
	\centering
	\includegraphics[width=\textwidth]{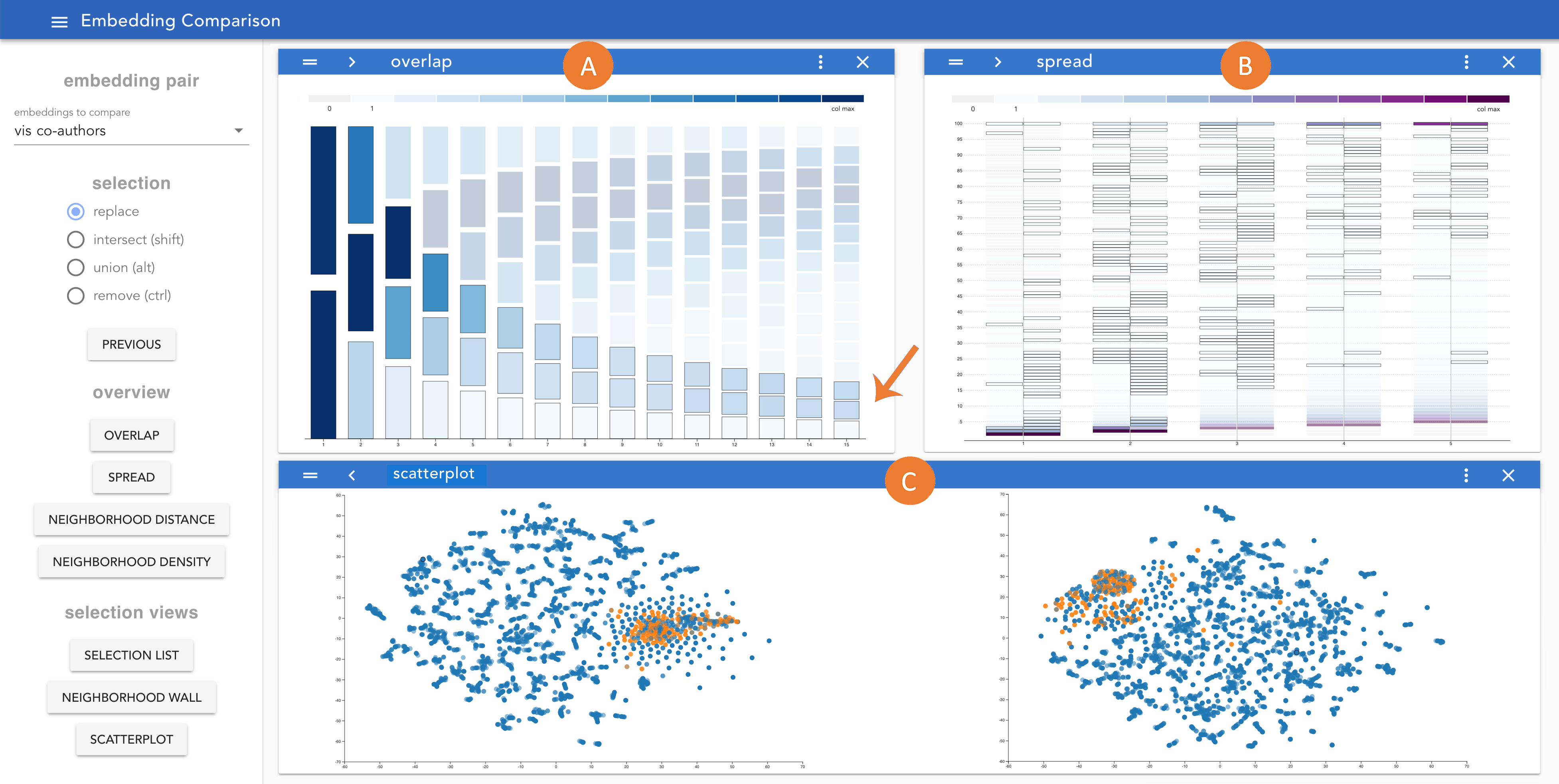}
	\caption{
		The picture shows the (A) \emph{neighborhood overlap}, (B) \emph{neighborhood spread} and (C) \emph{scatterplot} views  showing the VIS co-author graph embedding scenario discussed in Section 1.1.
		The user has clicked on the 3 boxes in the lower right of (A), marked by the orange arrow, to select all documents that share 2 or fewer of their 15 nearest neighbors.
	}
	\label{fig:motivation}
\end{figure*}

\subsection{Motivating Example}\label{sec:motivating_example}
We motivate our approach with an example where \sysname is used to compare embeddings generated by different algorithms.
The dataset is a publicly available collection of visualization publications~\cite{Isenberg2017} with all papers from the past 28 years to the VIS conference series.
Our goal is to use a co-author graph embedding for finding related authors. 
To build this embedding, we extract the co-author graph (with 5402 authors as its nodes, and co-author relationships as edges).
We then generate two different embeddings\footnote{Parameters for both representations: dimensions: 128, number of walks: 10, walk length: 80, window size: 10.} of the graph nodes, one with node2vec~\cite{Grover2016}, and the other one with deepwalk~\cite{Perozzi2014}.
We use \sysname to compare the embeddings (Figure~\ref{fig:motivation}) to understand the differences in how the methods model the data, and the potential effect of these differences in our application scenario.

To start, we use the \emph{overlap} view (Figure~\ref{fig:motivation}A) that provides a summary of the \emph{neighborhood overlap} metric. The metric shows how many author nodes have similar neighbors in the two embeddings. The details of this metric and the summary visualization are described in \S\ref{sec:pointwise}. In the visualization, the rightmost column shows the distribution of the metrics for the author nodes: we see that many nodes share 9-15 neighbors (dark areas at the top right), but some share 0-2 neighbors. Selecting the boxes corresponding to 0, 1, and 2 neighbors (lower right in Figure~\ref{fig:motivation}A) we select the authors whose neighborhoods have little similarity between embeddings. This causes them to be highlighted across all views currently open.

We want to understand the severity of the changes in distances that cause the differences in neighborhoods.
In other words, do neighbors in one embedding generally stay in the vicinity in the other embedding, or do they move further away?
For this, we use the \emph{spread} metric that is summarized by Figure~\ref{fig:motivation}B (details are described in \S\ref{sec:pointwise}). The view shows that overall, many of the nearest neighbors remain close between embeddings.
However, it also shows that the selected authors have neighbors that move farther away (shown by boxes with a black outline) in the upper half of the view, starting from column 3.

We continue to investigate by selecting subsets and viewing them in a number of additional, close-up views (details about these views are discussed in \S\ref{sec:system_design}).
Using the \emph{selection list} (not shown) we can examine the nodes with low overlap and see they are authors with a very low publication count and, therefore, little connection to the rest of the co-author graph.
Using the \emph{scatterplot view} (Figure~\ref{fig:motivation}C), we can see differences in the patterns of the selected nodes: while \emph{node2vec} places those nodes in a high-density area in the embedding very close to similar, unconnected, authors, \emph{deepwalk} tends to distribute them throughout the embedding next to unrelated authors.

Two additional metrics, \emph{distance} and \emph{density}, and their overview visualizations (details are discussed in \S\ref{sec:distribution}) help us identify unconnected authors in the \emph{node2vec} embedding.
It places them far away from the other nodes in a low-density region of the embedding.
We can use this knowledge to identify unconnected authors and use it to increase precision of our author suggestion function.
In this analysis, we identified an important difference between both embedding methods, which helps us decide which one we prefer for our application scenario.
Closer inspection further provided us with insights about how to use properties of the embeddings to extract additional information that is of relevance in the domain.

\section{Related Work and Background}\label{sec:relwork}
\label{sec:related_work_embeddings}
Embeddings are generated by inferring structure from input data, and represent the data objects as vectors.
Closeness in the vector space can be interpreted as high similarity between both data objects.
To compute and compare distances between object pairs in embeddings, we need a distance function.
Different embedding methods rely on different ways to compute distances between vectors in the resulting spaces.
Our approach is metric agnostic, and can be applied to any vector space for which the object distance can be computed.

Embedding algorithms are complex and often take high-dimensional data as input.
They then find a representation of that data in a lower dimensional space closer to the input data's latent space such that all relevant relationships between the data objects are retained, while noise is eliminated.
An example for this is the GloVe word embedding method~\cite{Pennington2014}.
It takes a matrix of word co-occurrences for every possible word pair as input, which usually has many hundred thousand dimensions.
The resulting vector space, that relates words to each other, typically has a few hundred dimensions.

Embedding methods exist for many different types of complex data objects.
All of them can be compared with \sysname.
Domains and data types include network~\cite{Perozzi2014}, medical~\cite{Choi2016}, biological~\cite{Yuan2017}, image~\cite{Hristov2018}, or citation~\cite{Berger2017} data.
A data type that relies on embeddings heavily is text data, for which topic models are a popular type of embedding method.
They abstract a document collection into coherent sets of words, called topics, and represent documents in a topic space.
Two common topic modeling methods are Latent Dirichlet Allocation (LDA)~\cite{Blei2003} and Non-negative Matrix Factorization (NMF)~\cite{Sra2006}.
Topic models can be interpreted as dimensionality reductions of a high-dimensional document-term matrix.
Dimensionality reductions are a particular type of embedding, where a high-dimensional representation of the data is reduced to a lower dimensional representation (embedding) that keeps some of the features of the original one.

Word embeddings place words into a vector space to encode semantic relationships.
Methods to create word embeddings include word2vec~\cite{Mikolov2013}, GloVe~\cite{Pennington2014}, and more recent improvements~\cite{Mikolov2017, peters2018}.
Recently, alignment methods for word embeddings have been proposed that create a mapping between two word spaces~\cite{Alaux2018, Grave2018}.
They serve specific linguistic tasks, but are not useful for comparing embeddings because they do not encode local differences between embeddings in an accessible way.
Those local differences, which our method helps uncover, are the most relevant properties to guide comparison.

Another type of embedding approach is dimensionality reduction, often used to visualize data.
They transform one embedding space into another, lower dimensional (often 2D) one, and try to retain distance and neighborhood relations as much as possible for visualization.
Popular methods include Multi-Dimensional Scaling (MDS)~\cite{kruskal1964multidimensional}, t-stochastic neighborhood embedding (t-SNE)~\cite{Maaten2008}, local linear embedding (LLE)~\cite{Roweis2000}, and Principal Component Analysis (PCA)~\cite{wold1987principal}.
Any of the previously discussed approaches can be used with our approach.

\subsection{Metrics for Dimensionality Reduction and Networks}\label{sec:metrics_for_dim_red}
Our approach is based on metrics that quantify embedding correspondences.
For the specific case of dimensionality reductions, metrics exist to gauge the quality of low dimensional representations.
A popular one is stress~\cite{kruskal1964multidimensional}, the sum of squared errors of pairwise distances between each pair of objects when projected down to a lower dimensional representation.
In the general embedding comparison case, relative distances between objects play a more important role than absolute ones.
For this reason, we focus on relative metrics, including neighborhood-based ones, instead.
The idea of comparing neighborhoods has been used in the past for evaluating dimensionality reductions~\cite{Lee2009, Mokbel2013}.
Correspondence metrics between neighborhoods in the original and the dimensionality-reduced version of a dataset include trustworthiness and preservation~\cite{Venna2001}.
They measure \emph{intrusions} and \emph{extrusions} into neighborhoods, respectively.
While those measures are not directly applicable in our approach, because they summarize neighborhood correspondences into one single number, some of our metrics (\S\ref{sec:metrics_and_overviews}) borrow from these ideas.

In addition to numeric quality metrics for dimensionality reductions, there are a number of visual interactive approaches that help users with interpreting and assessing reduction results.
For example, VisCoDeR~\cite{Cutura2018} allows comparisons between 2D dimensionality reduction methods by enabling brushing and linking between their respective results.
This type of comparison is not feasible for higher dimensional embeddings, because they cannot be easily represented visually.
Stahnke et al.~\cite{Stahnke2016} help users evaluate one specific 2D representation of a higher dimensional dataset by relating it to the object features in the original, higher dimensional space.
This requires interpretable features in the original data space, which often do not exist.
DimStiller~\cite{Ingram2010} guides users through the process of creating dimensionality reductions, while Dis-Function~\cite{Brown2012} allows users to interactively modify distance function in a 2D space.
These approaches focus on interpreting dimensionality reductions, while we consider more general comparison between embeddings.

Metrics that describe large and complex data structures have been developed and used for network data.
Bagrow et al.~\cite{Bagrow2008} introduced the B-matrix format for networks that captures distributions of neighborhood sizes of network nodes based on hop count.
They use a color ramp-based display to visualize and compare network structures.
GraphPrism~\cite{Kairam2012} implements a variation of the B-matrix visualization for large networks and combines it with a range of additional network metrics into a system for analyzing and comparing network structure.
Some of \sysname's metrics are based on ideas similar to that of a B-matrix, but target vector spaces with different properties and requirements than networks.

\subsection{Visual Model Creation and Comparison}
Object embeddings are data-driven models that abstract from complex data sets.
In this section, we review related visual approaches to model guidance and comparison.
While the goals, comparison targets, and challenges~\cite{Gleicher2018} of the approaches differ, they are all in some aspect related to this work.

Many approaches address the comparison of specific model types, often for specific usages.
For example,
Serendip~\cite{Alexander2016} provide tools for comparing topic models.
Clustervision~\cite{Kwon2018} helps
users compare clusters with different quality metrics,
while iVisClustering~\cite{Hanseung2012} allows users to interactively explore a topic model on a set of documents and interactively refine the model by splitting and combining topics.
TreePod~\cite{Muehlbacher2018} helps to interactively create decision tree models based on trade-offs between model properties.
These previous approaches focus on other models and tasks, while we aim at embedding model comparison for a broad range of scenarios.

Other recent approaches consider interpretation of word embeddings.
Some focus on analogy relationships between word vectors~\cite{Liu2018, Chen2018}, generate meaningful representations of the embedding space~\cite{Smilkov2016}, or help users with specific linguistic tasks, such as the creation of semantic word fields~\cite{Park2018}.
Heimerl and Gleicher focus on local phenomena~\cite{Heimerl2018}, while Parallax~\cite{Molino2019} allows users to view subsets of embeddings based on subspaces meaningful to the user.
Latent Space Cartography~\cite{Liu2019} is a method that allows users to map and compare multiple vector spaces based on user-defined example vectors.
None of the above support comparisons on a global scale between embeddings without requiring the user to have an initial hypothesis about similarities and differences.

EmbeddingVis~\cite{Li2018} introduces metrics for the network domain to compare nodes within embeddings to the source networks.
In contrast, we focus on comparing any kind of embedding and data types.
Parallel Embeddings~\cite{Arendt2020} uses dimensionality reductions and visualizations of corresponding clustering structure in embeddings to help users gauge the relative performance of classification models based on the embeddings compared.

\subsection{Design Inspirations}
We use color fields to encode and compare 1D distributions of our embedding metrics.
This basic design supports interaction with the data and is inspired by Piringer et al.~\cite{Piringer2012}, who use color fields to compare 2D distributions of data.
Another source of inspiration~\cite{Sopan2010} uses, among other designs, color fields to convey value distributions in tabular data.
Szafir et al.~\cite{Szafir2016ensemble} discuss the perceptual issues in using color fields for summaries.

Van Ham and Perer~\cite{vanHam2009} support navigation and drill down in large network structures to facilitate exploration based on a metric that quantifies potential interestingness of subgraphs.
Heimerl et al.~\cite{Heimerl2016} help navigate large document scatterplots and change focus to various levels of detail.
Neither of these approaches focuses on comparison tasks.

Our system uses multiple coordinated views that allow a user to configure the interface to their needs \cite{Roberts2007}.
We use flexible selection mechanisms to allow for exploration, an idea we further developed in the Boxer system for classifier comparison~\cite{Boxer}.

Prior systems have used embeddings to learn about the underlying data set.
Chuang et al.~\cite{chuang2012} uses topic models to facilitate exploration of document similarities and highlight potential inaccuracies of the model representation.
Cite2vec~\cite{Berger2017} embeds citations together with words to analyze how they get used in the citing literature.
We use elements from these systems in our design.

\section{Design Rationale: Scenarios and Challenges}\label{sec:tasks_and_requirements}
We apply a framework for comparison~\cite{Gleicher2018} to provide an abstraction of the embedding comparison problem, identify the key comparative challenges, and suggest solution strategies.
First, we consider a range of example scenarios and abstract common tasks and data properties.
Then, we identify comparative challenges and connect them to a solution strategy.

\subsection{Example Scenarios}\label{sec:example_scenarios}
We survey a set of representative scenarios of how embedding comparison is used in practice.
These examples illustrate the potential utility of an embedding comparison tool, common tasks, and the diversity of challenges.
We draw these examples from our experiences using embeddings, from collaborators interested in scenarios involving embedding comparison, as well as the literature.

\begin{enumerate}[label=\textbf{S\arabic*}]

\item \label{ex:hd2d}
A designer creates a 2D reduction of high-dimensional data to present it visually.
They must check that the visual presentation adequately conveys the contents of the high-dimensional space.
They first check the overall amount of local structure preserved to identify potential problems and then investigate specific objects to see if these errors are acceptable.
For this, simple summary metrics are not sufficient because they do not give users access to subsets of objects or single objects for assessment.
This is a common scenario~\cite{chuang2012, Koeper2016}.
While data projection methods attempt to retain pairwise distances~\cite{kruskal1964multidimensional} or cluster structure~\cite{Maaten2008}, quality assessment must be applied on a case-by-case basis.

\item \label{ex:hypp}
A researcher performs model selection between topic models with different hyper-parameters.
To identify the best model, they must make subjective assessments of which of these embeddings provides better similarity recommendations.
They first survey the differences across the whole embedding to assess the amount of difference and identify documents with different recommendations.
These differing neighborhoods are examined to understand differences and learn about potential causes.
Assessing document embeddings that help find similar documents and analyze similarity and cluster structure of a text collection is a frequent task (e.g., ~\cite{Alexander2016, Crossno2011}).

\item \label{ex:vecsz}
A developer may compare word vector embeddings built with different numbers of dimensions, algorithms, or hyper-parameters to understand the effects of these changes. To assess the differences in results, they may see how often words have different neighborhoods and what words are most affected.
Users of embedding methods often perform such tests using standard evaluation measures~\cite{Heimerl2018} of single metrics.
There are no existing tools to systematically compare such spaces.
This is a standard scenario in the word embedding literature~\cite{Pennington2014, Mikolov2013}, and similar challenges exist for other data objects, such as images~\cite{Hinton2006} and graphs~\cite{Grover2016, Perozzi2014}.

\item \label{ex:overfit}
A data scientist compares two supervised dimensionality reductions built from different amounts of training data to analyze potential overfitting of the learning algorithms.
After confirming the overall difference between the embeddings, they can identify specific objects to explore the relationships with the training data.
Model comparison scenarios for supervised dimensionality reduction are described in the literature~\cite{Etemad1997, Zhong2013}.

\item \label{ex:corpora}
A literature scholar compares word vector embeddings, each built from a different collection of texts to analyze the differences in word usage.
This involves identifying words that stay constant or have dramatic changes, examining specific words to see how their neighborhoods change, and reconnecting these findings to the overall language by looking for groups and patterns of words that behave similarly.
See Section~\ref{sec:eebo} for an example.
Similar scenarios are also described in the literature~\cite{Hamilton2016, Jatowt2014}.

\item \label{ex:languages}
A linguist may compare word embeddings constructed from corpora in different languages in order to understand latent differences in word usage between the languages.
Here, the objects (words) in the different embeddings are not the same; however, there is a known correspondence between some of them.
This scenario has precedence in the literature~\cite{Thompson2018}.

\end{enumerate}

Users currently address these kinds of tasks using either \emph{ad hoc} analysis or specialized tools.
We use the list to abstract some key ideas that inspire a general tool for embedding comparison.

\noindent\textbf{Global and local tasks:} Each scenario follows a similar pattern of tasks:
(1) generally assess the amount of difference between the embeddings,
(2) identify a set of objects whose neighborhoods differ in interesting and relevant ways,
(3) more closely assess the objects and their neighborhoods.
The changes in neighborhood may then be dissected to determine their cause, or re-connected to the overall embedding by finding other objects with similar behavior.
This list provides a core set of abstract tasks that our approach must address.
Using the target/action task notation~\cite{Munzner2014} as adapted for comparison~\cite{Gleicher2018}, there are several common comparison targets: entire embeddings, collections of objects, and the neighborhoods of objects.
Important actions include \emph{quantifying} the amount of difference, \emph{identifying} interesting sets and objects, \emph{dissecting specific differences to understand their cause}, and \emph{re-connecting} specific findings to the larger set or their impact on downstream uses.

The ability to identify and analyze the changes in neighborhoods of objects is central to all comparison scenarios we have considered.
Differences in an object's set of neighbors reflect differences in the type of relationships encoded between objects.
This ability is particularly important in the common cases where there is no ground truth, so assessment requires some degree of subjectivity.
For example, if a document has very different neighbors in two topic models, an expert may want to evaluate if one set is preferable to the other, or whether the document can be disregarded as an outlier.
Identifying specific objects is often more interpretable to a user than more abstract relationships between objects, and connects to the semantics of the embedding.
For example, in a word embedding, the user most likely understands the objects (words) and has considerable background knowledge to interpret how words behave.

\noindent\textbf{Understanding models vs. understanding data:}
Users have two types of goals.
Some scenarios (\exhdtored, \exhypp, \exvecsz) involve comparing different embeddings of  the same data in order to understand differences between the models.
In contrast, other scenarios (\excorpora, \exlanguages) involve comparing embeddings built from different data to compare the underlying data sets.

\noindent\textbf{Different uses of embeddings:}
The scenarios all identify similarities and differences between neighborhoods in embeddings, but use them in a variety of ways.
Some scenarios seek to identify similar objects (\exhypp), while others use this information for other processes such as clustering or nearest-neighbor classification (\exoverfit), or creating 2D models for visual presentation that place similar objects close together (\exhdtored).
All of these uses of embeddings focus on the similarity information in the embedding; the points assigned to objects are less important than the distances between objects.
In many cases, such as word vector embeddings or t-SNE dimensionality reduction, the embedding spaces are abstract, i.e., the dimensions have no meaning other than providing a space where relationships between objects hold.
In some cases, such as topic models, the dimensions may have some semantic meaning.
For such scenarios, global positions may be important.
These can be addressed with standard high-dimensional analysis techniques~\cite{Liu2017}.

\noindent\textbf{Differences in embedding spaces:}
All scenarios may involve comparing embeddings in different spaces.
The spaces may differ by numbers of dimensions, or have very different scalings.
Such differences preclude direct comparison of object coordinates, or even of the distances between objects.
Therefore, we focus on using distance ranks (i.e., orderings of closeness) as such measurements are invariant to dimensionality and scaling; two objects can be nearest neighbors in very different spaces.
However, distance values may be important in understanding the causes of distance rankings.
For example, a small change in an object's distance may cause a large change in its rank if there are many objects with similar distances.
This means we must consider both rank- and distance- based metrics and allow for using the two types together.

\subsection{Challenges and Solution Strategy}
\label{sec:strategies}
Gleicher~\cite{Gleicher2018} identifies three categories of comparative challenges.
We focus on pairwise comparison, avoiding the challenge of large numbers of targets for comparison.
This leaves the challenges of large targets (embeddings may have large numbers of objects) and complex relationships (each object's local neighborhood may change in many ways).

\noindent\textbf{Challenges of Complexity:}
The complexity of embedding comparison stems from the fact that the important similarities, e.g., whether objects are close together, may be obscured by other differences, e.g., their positions in space.
To manage this complexity, we focus our attention on the objects and their neighborhoods, including the distances between objects in these neighborhoods.
We break the comparison of embeddings into the comparison of the neighborhoods of each object in the embedding.
Given a pair of neighborhoods corresponding to an object in a pair of embeddings, we can define a number of metrics that capture the degree of difference in the distance relationships.
More global comparisons consider the distributions of these per-object comparisons.
\S\ref{sec:metrics_and_overviews} considers a set of metrics for comparing neighborhoods, and introduces a view for summarizing distributions of each metric for global assessment.
In addition, detail views help with examining specific objects and their local neighborhoods.

\noindent\textbf{Challenges of Scale:} To address the challenges of scale, we use all three scalability strategies, \emph{summarize}, \emph{select subset}, and \emph{scan sequentially}~\cite{Gleicher2018,alpersummarize}.
First, we use \emph{summary} views that aggregate per-neighborhood metrics across the embeddings.
These views show the distributions of the various metrics.
They allow the user to both have an overall sense of the amount of similarities and differences, and help identify sets of objects with interesting values of the metrics for further exploration.
These interesting subsets may be \emph{selected} to focus on a more manageable portion of the object collection.
Our approach provides mechanisms for creating complex selections by combining groups seen in summary views (\S\ref{sec:metrics_and_overviews}), and provides views that allow details of moderate sized groups to be examined (\S\ref{sec:selection_details}).
\emph{Scanning} in these intermediary views is used to identify specific objects to examine details.

\subsection{Running Example Use Case}\label{sec:example0}
We will use one use case as a running example throughout the next two sections to motivate the methods introduced and illustrate what they can be used or in practice.
Here, we introduce the use case and will return to it at the end of both Sections~\ref{sec:metrics_and_overviews} and~\ref{sec:system_design}.
In this example, we want to assess the quality of a dimensionality reduction by comparing a bag-of-words embedding (every word in the collection represents a dimension) of documents to its dimensionality-reduced version.
This is an example of scenario \textbf{\ref{ex:hd2d}}.
The dimensionality-reduced space is two-dimensional, therefpre, it can be used for presentation and visual analysis of the source data.

This example uses t-SNE~\cite{vanHam2009}, a dimensionality reduction method for 2D projections that is popular because it aims to conserve the neighborhoods from the original space.

Traditionally, metrics are used to quantify the quality of dimensionality reductions.
The most popular one is stress~\cite{kruskal1964multidimensional}.
Stress and all other metrics discussed in \S\ref{sec:metrics_for_dim_red} quantify quality as one value.
This keeps all details and advantages and disadvantages of a 2D representation of the data hidden.
Often, users want additional information about how good and representative the dimensionality-reduced version of the dataset is by inspecting concrete examples and their surroundings.
To support this, our approach treats dimensionality reduction as a comparison of two embeddings, the original and the dimensionality-reduced one.

\section{Comparison Metrics and Overviews}\label{sec:metrics_and_overviews}

\begin{figure}
	\centering
	\includegraphics[width=\columnwidth]{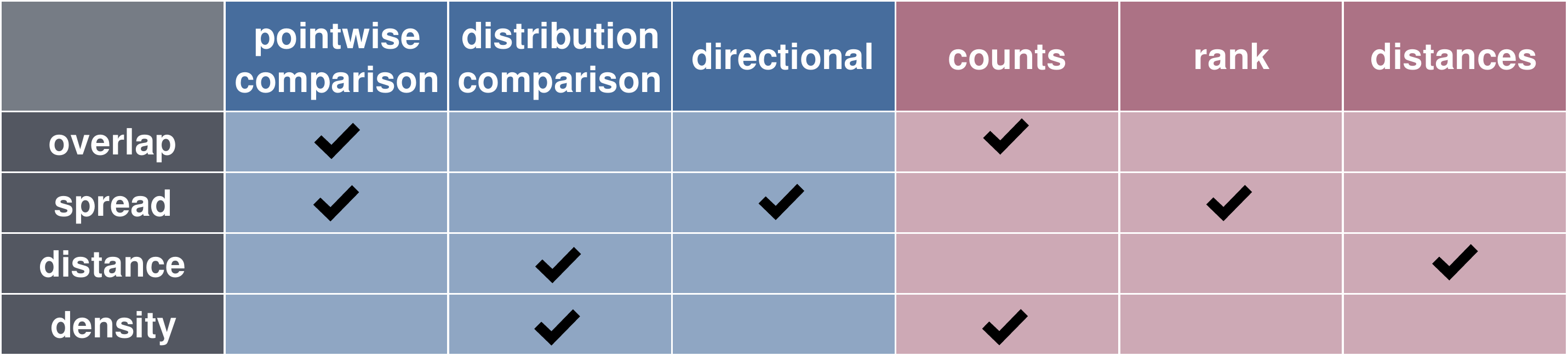}
	\caption{
		The differences between the overview visualizations (\emph{overlap}, \emph{spread}, \emph{distance}, and \emph{density}), including \textbf{characteristics of the metrics and designs}~\crule[tableleft]{.3cm}{.3cm} and \textbf{data types for the distributions}~\crule[tableright]{.3cm}{.3cm}.
		Characteristics of the metric determine the design variant used to convey per-object distributions of the metrics across an embedding pair.
		To cover all relevant characteristics, we provide four variants, as shown in Figure~\ref{fig:compilation}.
		Data types are the types of measurements we use within or between embeddings.
		They are important for interpreting the visualization, and determining whether the per-object distribution is discrete or continuous.
		In the latter case, distributions are binned before they are visualized.
	}
	\label{tab:overview_table}
\end{figure}

Our approach is based on four core metrics that address different aspects of embeddings.
Each metric is based on concepts from the literature, and has been adapted for the specific embedding comparison scenarios we support.
For each metric, we provide a visualization to convey an overview of the distribution of
the metric values for each item across the embedding pair.
Those summary visualizations provide a synoptic perspective onto similarities and differences.
Based on these views, users can identify potentially interesting objects and select them for closer inspection.

We have found that the four metrics cover the range of tasks and scenarios discussed in \S\ref{sec:tasks_and_requirements}.
Over multiple iterations of design and evaluation, we found that this set is sufficient.
Figure~\ref{tab:overview_table} shows a summary of the four metrics
and their characteristics and corresponding overview visualizations.
Views with the \emph{pointwise comparison} characteristic show distributions of metrics that directly quantify differences between local object neighborhoods in embeddings.
Views with the \emph{distribution comparison} characteristic show metric distributions per embedding, juxtaposing distributions of the two embeddings.
Comparison takes place visually between the juxtaposed distributions.
\emph{Directional} views are based on metrics for which the order of the embeddings plays a role, so they yield two distributions per embedding pair.
The possible data types that the metrics yield distributions over are \emph{counts}, which are numbers of objects, \emph{rank}, which are rank positions in an object neighborhood, and \emph{distances}, which are distances between objects in an embedding.

\begin{figure*}
	\centering
	\includegraphics[width=\textwidth]{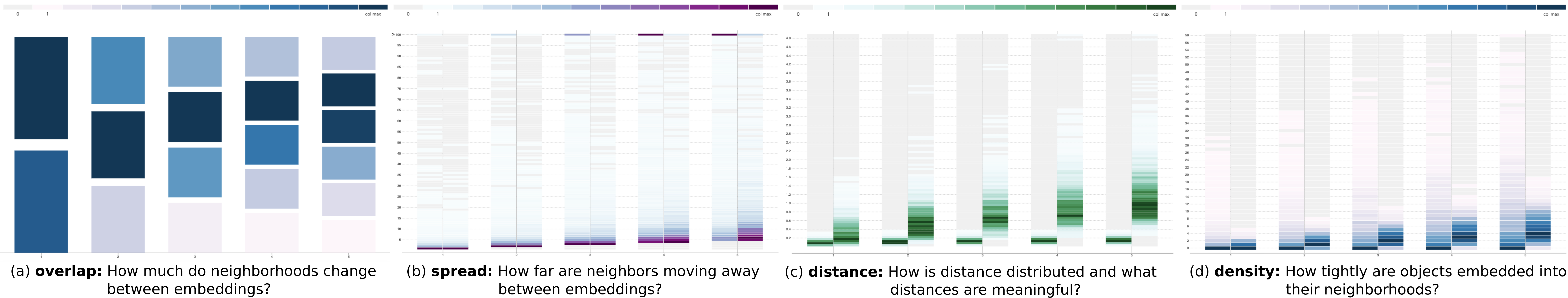}
	\caption{
		All four variations of the distribution comparison visualization, each one based on a different metric to compare embeddings.
		We use different colors based on the data type of the distributions for \textbf{counts}~\crule[countscolor]{.3cm}{.3cm}, \textbf{ranks}~\crule[rankscolor]{.3cm}{.3cm}, and \textbf{distances}~\crule[distancescolor]{.3cm}{.3cm}.
		Those overview visualizations serve as input and output.
		Each of the bins of the distributions can be selected by users, highlighting all other bins that contain any of the selected objects.
	}
	\label{fig:compilation}
\end{figure*}

For each metric, we provide visual summaries through variations of the same design.
This helps users compare distributions for each of our metrics.
Figure~\ref{fig:compilation} shows them next to each other.
All of them are based on a color ramp display to convey distributions.
Similar visual encodings for distributions have been proposed previously~\cite{Sopan2010}.
We use ramps based on different colors to help distinguish between the data type of the distributions.
Those are \textbf{counts}~\crule[countscolor]{.3cm}{.3cm}, \textbf{ranks}~\crule[rankscolor]{.3cm}{.3cm}, and \textbf{distances}~\crule[distancescolor]{.3cm}{.3cm}.
While we tested alternatives, including bar charts and violin plots, we decided to use color ramp displays as our default encoding.
Color ramps make it easy to distinguish between zero and very low values by using a different color for zero.
This is important for outliers and long tails in the distributions.
Furthermore, each segment of the display has a constant shape, enabling interaction with even small segments for selection.

The color ramps in Figure~\ref{fig:compilation} (legends are on top of each of the plots), map count values to colors.
Zero is encoded by the distinctive color gray.
All counts denote the number of objects for which the respective metric yields the corresponding value along the vertical axis.
Distributions are organized as columns in the plots, with low to high values from bottom to top.
The variation in Figure~\ref{fig:compilation}a varies in the number of bins along the vertical axis, the other three views have a constant number of bins for each column.
The variations in Figure~\ref{fig:compilation}b-d show two distributions immediately next to each other for easier comparison between them.
We now introduce the metrics, and connect each one to its specific design variation.

In the following, we introduce our notation.
Local neighbors are other objects in an embedding that are closest to a given object.
Let $d_{e}(i,j) \geq 0$ is the distance between two objects $i$ and $j$ in embedding $e$.
Every embedding can have a different distance metric associated with it.
In practice, we typically use either Euclidean or cosine distance.
The list $s_0^e(i), s_1^e(i), s_2^e(i), ..., s_n^e(i)$ contains all elements sorted, smallest to largest, by their distance to $i$ in $e$.
It is, in other words, the ordered list of nearest neighbors of object $i$ in embedding $e$.
The n-neighborhood of an object $i$ in embedding $e$ is $\mathrm{neighborhood}_{e,n}(i) = \{ s_0^e(i), ..., s_n^e(i) \}$, the set of the n closest neighbors to $i$ in $e$.
An item $j$ has rank k, $rank_e^i(j) = k$, if and only if $j=s_k^e(i)$ in embedding $e$.
This means that the position of item $j$ in the neighborhood list of object $i$ in embedding $e$ is k.

\subsection{Pointwise Comparison Metrics}\label{sec:pointwise}
Pointwise views are based on metrics that directly quantify local differences in object neighborhoods between embeddings.
They are derived from metrics of neighborhood correspondence, which have been used in the literature to measure the quality of dimensionality reductions~\cite{Lee2009, Mokbel2013}.
Correspondence metrics include trustworthiness and preservation~\cite{Venna2001}, which measure \emph{intrusions} and \emph{extrusions} into neighborhoods.

\noindent \textbf{Overlap Metric:}
The overlap metric measures the number of common objects in both neighborhoods of an object.
For a given neighborhood size and object, it yields one value per embedding pair that denotes the number of identical neighbors of that object in both embeddings for the given neighborhood size.
Overlap is a symmetric metric between neighborhoods of two embeddings, i.e. the order of the embeddings does not play a role for the results.
It is defined as:
$$overlap_{e_1,e_2}(n, i) = | \mathrm{neighborhood}_{e_1,n}(i) \cap \mathrm{neighborhood}_{e_2,n}(i) |$$

A summary of the overlap metric is shown in the overlap view (Figure~\ref{fig:compilation}a) .

The columns of this view convey distributions of counts over the objects ($i$).
The horizontal axis of Figure~\ref{fig:compilation}a shows values
different
neighborhood sizes ($n$).
For size 1, the distribution has two bins, representing the two possibilities for neighborhoods of size one: there can be an overlap of either 0 (different nearest neighbor, bin on the bottom) or 1 (same nearest neighbor, bin on the top).
The number of possibilities (bins) increases with increasing neighborhood size.
A color encoding indicates the number of items in the bin.

\noindent \textbf{Spread Metric:}
The previous metric helps to find neighborhoods that change through analyzing its distributions across an embedding pair.
A common next question is where those changing neighbors move to.
This is where the rank-based \emph{spread} metric can help.
The spread metric yields the maximum neighborhood rank that any member of the neighborhood of object $i$ of size $n$ in embedding $e_1$ has in embedding $e_2$:
\[
s_{e_1,e_2}(n, i) = max( rank_{e_2}^i(s_1^{e_1}(i)), \ldots, rank_{e_2}^i(s_n^{e_1}(i)) )
\]
Spread can be interpreted as how far a neighborhood stretches or extends from one embedding to the next.
It can be viewed as a summary of the co-ranks within a co-rank matrix~\cite{Lee2009, Mokbel2013} that only includes the maximums.
For this metric, the order of the embedding plays a role.
It is thus marked as directional in Figure~\ref{tab:overview_table}.

Figure~\ref{fig:compilation}b shows the visual encoding for the spread metric.
Neighborhood sizes are plotted along the horizontal axis.
Each of the columns has two sides with two different distributions.
These are the distributions for both directions of the metric.
In this example, we can see that the nearest neighbor (neighborhoods of size 1 in the first column from the left) generally stays close for the example embedding pair because of the dark purple colors close to the bottom of the distribution in the first column of Figure~\ref{fig:compilation}b.
However, there are also a significant number of outliers (the light blue bins on the upper end of the distribution).
Spread yields a rank for each ordered embedding pair, object, and neighborhood size, as shown in Figure~\ref{tab:overview_table}.
The upper limit of the distribution scale is 100, which we found to work well as a default value.
It can be adapted freely by users.
Thus, the resulting distributions are distributions over a discrete domain (100 bins by default along the vertical axis).

\subsection{Distribution Comparison Metrics}\label{sec:distribution}
Distribution comparison metrics quantify local properties of objects within single embeddings, while the previous metrics quantify differences between embeddings directly.
Comparison between embeddings happens by comparing the distributions of a metric for each of the embeddings.
Both design variations for distributions of these metrics use two-sided columns (see Figure~\ref{fig:compilation}c and d).
Each of the sides shows the distribution for a different embedding.
This makes comparisons between distributions straightforward.

Both distribution comparison metrics consider the local configurations around an object.
Analyzing their distributions gives users insights about the differences in the placement of objects and their neighbors in both embeddings.

\noindent \textbf{Neighbor Distance Metric:}
This metric quantifies raw distances to the nearest neighbors in both embeddings.
For each object $i$, embedding $e$, and neighbor rank $k$, the neighbor distance metric yields the distance between the object and its k-th neighbor: $d_e(i, s_k^e(i))$.
This distance is shown along the vertical axis of Figure~\ref{fig:compilation}c.
In every column, Figure~\ref{fig:compilation}c shows two distributions, one for each embedding.
Users can specify the visible distributions through the range of neighbor positions.

In contrast to the previous metrics, the \emph{neighbor distance} metric yields distributions over a continuous variable (distances in the metric defined for the embeddings).
For this reason, the domain of the distribution is discretized before it can be displayed.
By default, we found that 100 bins is a good resolution for this view, but users can freely increase or decrease resolution if needed.
Direct comparisons between distances in the embeddings, especially with different dimensionality, may not be meaningful.
The views of this metric help to compare distances by providing context for the entire embedding.
For example, an object can be an outlier in one embedding, far off from its next neighbors, but right at the second embedding's distribution center.
We can see in Figure~\ref{fig:compilation}c that distances are smaller overall in the embedding on the left (the entire range of the distribution stays pretty low on the vertical axis).

\noindent \textbf{Local Density Metric:}
This metric quantifies densities around objects in each of the embeddings.
In order to keep both distributions comparable across the embeddings, we measure density as the number of neighbors within certain radii around an object in an embedding.
At first, we compute the average distance $r^{e}_{k}$ to the k-nearest neighbors across all objects in embedding $e$, with

\[
r^{e}_{k} = \frac{1}{|e|}\sum_{i\in e}d_e(i, s_k^{e}(i)).
\]

Then, we use this distance as the radius around objects, and count, for each object $i$, the number of neighbors within $r^{e}_{k}$.
The more neighbors, the denser the local neighborhood of object $i$.
Embedding-specific determination of the radii scales the densities by the embedding average, making them comparable across embeddings.
The \emph{local density} metric yields count numbers of objects for each object and embedding pair, as indicated in Figure~\ref{tab:overview_table}.
This is a distribution over a discrete domain.

Figure~\ref{fig:compilation}d shows an example of the view for the \emph{local density} metric.
It shows distributions for different values of k (along the horizontal axis), that can be chosen by users.
For each column, one distribution per embedding is depicted.
By default, the maximum value displayed along the vertical axis is 100, which we found to work well in practice.
Users can adapt this value as needed.
Figure~\ref{fig:compilation}d shows us that density in the right embedding is generally higher than in the left embedding.
We can see this through the dark color spread higher up in the columns, which means that there are many objects that have neighbors in the respective radius.
The left embedding, however, has a few high-density outliers along the range of the scale (light blue and white bins higher up).

\subsection{Example Use Case}\label{sec:example1}
We now look at how we can use the metrics and their overview visualizations to compare a dimensionality reduction to the high dimensional embedding (scenario \textbf{\ref{ex:hd2d}} in \S\ref{sec:example_scenarios}).
The first thing we want to know is how local surroundings of objects change.
To get a first impression of this, we view distributions of the \emph{neighborhood overlap} metric.
It shows us, for each document, the number of neighbors that are identical in both its neighborhoods.
From Figure~\ref{fig:compilation}a, we can see that most documents have the same nearest neighbor (darker blue on the top bin in the first column), but some have a different nearest neighbor.
Larger neighborhoods generally have a high overlap (blue colors towards the top of the plot).
However, there are also a significant number of outliers with low overlap in their nearest neighbors across both embeddings (light blue colors on the bottom of the plot).
The \emph{neighborhood spread} view (Figure~\ref{fig:compilation}b) provides additional insights about where neighbors move that are not in the overlap sets.
Dark purple on the bottom of the plot indicates the neighbors that stay close across both embeddings.
In addition, we can see a range of outliers (light blue bins higher up) of neighbors that move further away.
In particular, there is a larger number (darker colors) high up on the left column.
This tells us that a number of neighbors that are close in the original space gets pulled apart on the lower dimensional one.
Further inspection through set selection and additional views can help us understand more about those examples (see \S\ref{sec:system_design}) for a continuation of this use case.

The distribution of neighborhood distances in Figure~\ref{fig:compilation}c shows huge differences in the variance of the distribution.
For both embeddings, we measure distances in Euclidean distance.
While the higher dimensional space (left columns) generally keeps neighbors very close, the 2-dimensional one places some documents far apart from their closest neighbors (light blue bins high up in the right columns).
The difference in variance is expected, because a higher number of dimensions provides more possibilities to position documents close to each other, while options are more limited in lower dimensional spaces.
The distribution of local densities (Figure~\ref{fig:compilation}d) also shows significant differences between embeddings.
While the high-dimensional one (left column) has a low density (dark blue at the bottom of the plot), there are a range of outliers in high-density areas (lighter bins higher up in the plot).
Through selection and closer inspection (the mechanisms for this are discussed in \S\ref{sec:system_design}), we can find out if those high-density documents and their close local clusters are accurately represented in the lower-dimensional representation.

\section{Visual Interactive Embedding Comparison}\label{sec:system_design}
The previously described overviews build the core of our approach for embedding comparison.
These views allow the user to select items of interest by selecting bins.
The views are complemented by 6 additional views and other functionalities that form a comprehensive interface.
These additional views help users analyze the details of selections, revise sets of selected objects by adding or removing objects, and inspect local properties of single objects.
Figure~\ref{fig:teaser} shows an overview of the system, with some of the views activated.
The views are grouped into three different categories: overviews, selection detail views, and local views.
Users can configure the desktop with different summary visualizations with selection detail and local views as needed for their analysis.
In addition, we have added functionality to provide some guidance to users with selecting and managing views based on the comparison scenario at hand (see \S\ref{sec:view_selection}).

\subsection{Selection Sets}\label{sec:selections}
Overviews help to identify subsets of objects that differ in interesting aspects.
Once users identify such a set, \emph{selection sets} help them isolating its objects and inspect them in greater detail.
\emph{Selection sets} are a subset of the objects representing a group of interest.
Users can define selection sets, for example, by selecting bins in overviews.
Users can modify selection sets using set arithmetic to combine them, for example, to intersect sets to create conjunctive queries, or to remove a particular set of objects.
Objects in the current selection set are highlighted across all views, letting users connect from a subset of instances to the global context through the overviews.
This helps identify potential causes for why the objects are modeled differently by the embeddings, but it also helps understand the diversity of the examples, which can lead to the discovery of additional examples of interest.

At any time, the system maintains a history of prior selection sets, allowing a user to return to previous work or create complex combinations.
To combine the active selection set with a new one using set arithmetic, users can change the selection mode in the \emph{configuration drawer} or by holding down the associated modifier key (see Figure~\ref{fig:teaser}A).

\subsection{Selection Detail Views}\label{sec:selection_details}
Selection detail views enable the user to focus on the selected set, scan it for interesting properties or interesting objects, and alter what is selected based on new insights.
All three selection detail views show representations of each object in the selection set.
The user can highlight these objects by hovering over them with the mouse.
Those highlighted objects are then displayed in the \emph{local views}, and highlighted in all activated \emph{selection detail views} to analyze them in detail.
In addition, highlighted objects trigger a tool tip that shows the meta data of the highlighted object.

\subsubsection{Selection List}
The selection list view shows a tabular representation of the objects in our selection set, as shown in Figure~\ref{fig:teaser}D.
Users can sort the list by meta data attributes, and they can use the search box to search for specific strings or numerical values to identify objects.
To modify the selection set, users can select rows and reduce the selection to them, clear the entire selection, or select the entire dataset.
With the last two functions, we are able to build selection sets in a top down or bottom up fashion by either starting with an empty set and adding specific objects to it through selection in the other views, or start with the entire dataset and filter specific objects based on their meta data or by de-selecting them in other views (using set arithmetic).

\begin{figure}
	\centering
	\includegraphics[width=.75\columnwidth]{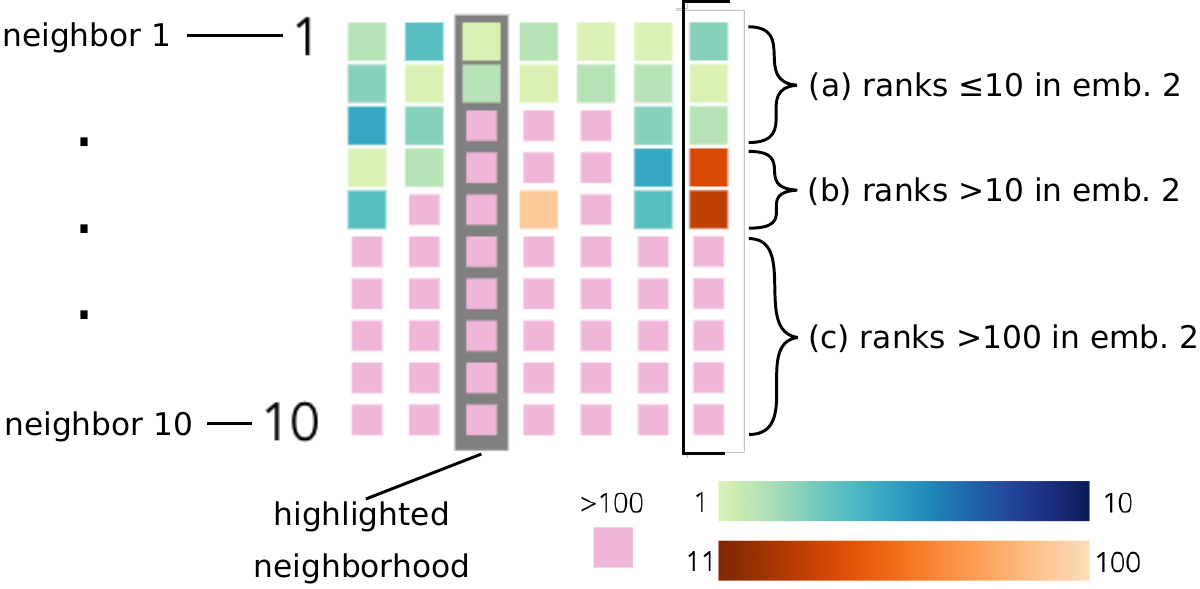}
	\caption{The encoding of neighborhood correspondences in the \emph{neighborhood wall view}.
				  Position from top to bottom encode neighborhood rank in one embedding, while colors encode rank in the second embedding.
			  	  Three different color scales tell us whether neighbors stayed within the neighborhood (a), moved out of the neighborhood, but stayed in the vicinity (b), or moved further away (c).
				  The ordering of colors relative to the color ramp conveys the change in ordering of the neighbors.
	  	  	  	  Neighborhoods can be sorted by a range of metrics, including overlap (number of yellow-blue boxes), and average change in rank. }
	\label{fig:wall}
\end{figure}

\subsubsection{Neighborhood Wall}
The neighborhood wall view, introduced in Figure~\ref{fig:wall}, provides an overview of the local correspondences between the neighborhoods of the objects in our selection set.
It complements the selection list view in that it focuses on the local neighborhood of the selected objects.
Users can get a sense of the diversity of changes in neighborhoods across embeddings in the selection set.
They can then, for example, select outliers or particularly interesting examples to further narrow down the selection, or to view them in more detail.

\begin{figure*}[ht]
	\centering
	\includegraphics[width=\textwidth]{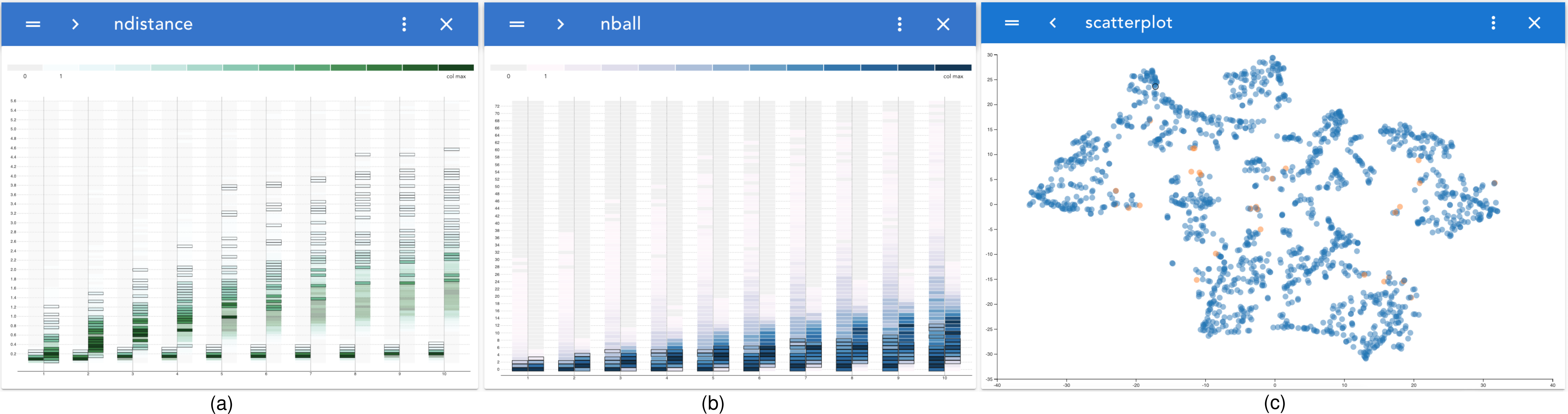}
	\caption{ \label{fig:big_picture} Three views that show the selection (28 objects) from Figure~\ref{fig:teaser}.
				   The \emph{neighbor distance view} (a) shows the distribution of neighborhood distances in the high dimensional (left) and the 2-dimensional (right) embedding.
		    	   The distance distribution on the left is much more widely distributed than in the right embedding.
		 		   In addition, we can see that the selected (black outlines) objects' neighbors are all relatively far away after the 5th neighbor in the embedding on the right.
				   The \emph{density view} (b) shows the densities around objects in both embeddings.
				   We can see high-density outliers (light blue bins) on the top of the distribution for high dimensional (left) embedding.
				   The \emph{scatterplot} (c) shows a scatterplot of the 2D embedding.
				   The dots are 50\% transparent, such that dark colors indicate that there is a lot of overlap of points and very tight clusters among the selected and the unselected objects.
			   	   Orange dots are part of the selection set. }
\end{figure*}

\subsubsection{Scatterplots}
Scatterplots based on dimensionality reductions of embeddings are commonly used to analyze and debug embeddings (e.g., \cite{Smilkov2016}).
\sysname provides a scatterplot view that can show either embedding, or the pair juxtaposed.
The scatterplot view is highly configurable to best serve the comparison problem at hand.
For embeddings with $>2$ dimensions, the view offers a range of dimensionality reduction methods (t-SNE, UMAP, LLE, PCA, MDS, and IsoMap).
Scatterplots of dimensionality reductions are potentially inaccurate, but are often useful in combination with the other views to corroborate hypotheses and confirm findings.

For very large and potentially dense embeddings, the scatterplot has a binning feature~\cite{Heimerl2018b} that provides an abstraction so that plots remain legible.
Figure~\ref{fig:uc2}c shows an example of a binned scatterplot.
In this particular design (which is adaptable by users), hexagonal bins are used in combination with pie charts that convey the ratio of selected (orange) and unselected (blue) objects in a bin.
Generally, in the scatterplot, objects in the active selection set are colored in orange.
By default, t-SNE with pre-defined parameters is selected to create the 2D representation.
The scatterplot allows us to change or modify our selection set by selecting single objects directly, or larger areas (e.g., clusters) using a rectangular selection tool.
This is possible in the binned and unbinned version of the scatterplot view.

\subsection{Local Views}\label{sec:local_views}
Local views allow the user to focus on one particular object in our selection set and compare it across embeddings.
This helps identify and analyze objects with interesting properties within or across the embeddings and relate them back to the global context.

\subsubsection{Neighborhood List}
The neighborhood list is a view that contains two lists similar to the one from the \emph{selection list view} (Figure~\ref{fig:teaser}D).
Each list contains the nearest neighbors in one embedding of the highlighted object ordered by the distance from the object.
In addition to object meta data, it also shows the distance to the highlighted object for each of the neighbors.

\begin{figure}
	\centering
	\includegraphics[width=0.9\columnwidth]{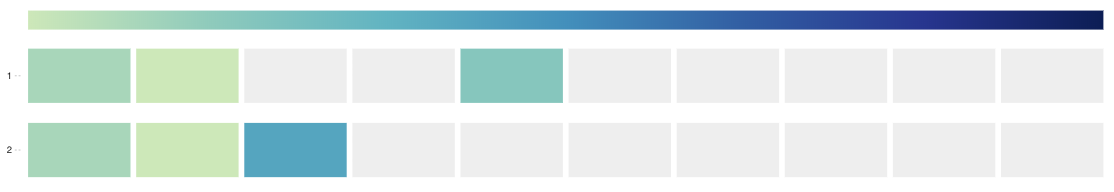}
	\caption{ The \emph{neighborhood sequence view} shows the neighbors for a highlighted object as colored boxes.
					Their position from the left encodes neighbors ordered by distance for the first embedding (top) and the second embedding (bottom).
					The color of the bars encodes the position of the neighbor in the second embedding based on the color ramp on top of the view.
					This is a variation of a \emph{buddy plot}~\cite{Alexander2016}.
					Gray boxes are neighbors that are not part of the neighborhood in the second embedding.
					We see that for 10 neighbors, 7 are different and the remaining 3 are ordered differently (colors are not ordered according to ramp).
					If users are interested in absolute distances, they can switch to distance mode (depicted in Figure~\ref{fig:teaser}F), which represents neighbors as bars and positions them according to their distances to the selected object. }
	\label{fig:buddyplot}
\end{figure}

\subsubsection{Neighborhood Sequence}\label{sec:neighborhood_sequence}
Compared to the \emph{neighborhood list view}, which allows us to scan and compare neighbors based on their meta data, the neighborhood sequence view focuses on comparing either the correspondence of ranks or the underlying distances within the neighborhood.
Figure~\ref{fig:buddyplot} explains the encoding.

\subsection{View Selection and Management}\label{sec:view_selection}
All of the views described are part of a prototype system (see \S\ref{sec:implementation} for details).
The current prototype's design provides freedom to users with respect to choosing and organizing views for comparing embeddings.
This makes the implementation flexible with respect to the scenarios that can be supported by it.
On the other hand, it also means that before being able to use it effectively, the user has to decide which of the views are likely going to be central to the comparison tasks at hand, and put in effort to create an initial layout of views that supports to scenario well.

In practice, there are patterns of choices of views and layouts that occur frequently.
In particular, we found the following three configurations to be most useful across a range of scenarios:
\begin{itemize}
\item The \emph{overlap} view is at the top of the desktop, with the \emph{selection list} and \emph{neighborhood wall} views next to it.
This configuration is the one we use most often.
It helps us find similarities and differences quickly by comparing neighborhood overlap, and we can start drilling down by selecting them in the overlap views, and inspecting them more closely in the selection detail views.
Other views, such as additional overviews and local detail views can be added below as needed.
\item The \emph{selection list} view is at the top of the desktop, and the \emph{overlap} and \emph{spread} views are next to it.
This configuration is useful when users have a good idea about what objects they are most interested in, for example, when comparing word usage through word embeddings.
The selection list can be used to select specific words, and the overviews provide context from the embeddings (e.g., about how constant their neighborhood is or how far away neighbors move).
Additional relevant local views can be added further down on the desktop.
\item The \emph{scatterplot} view is at the top of the desktop, with the \emph{overlap}, \emph{selection list}, and \emph{neighborhood wall} view around it.
This configuration is useful when both embeddings are 2-dimensional, and their quality and similarity is to be analyzed.
The scatterplot can then be used to select potentially interesting groups of examples (e.g., a cluster), and the other views provide global as well as local context for the objects.
\end{itemize}

While these configurations are not exhaustive, they provide an entry point to the functionality of the prototype, and can be modified and adapted if needed.
We have added options to start with any of these configurations to speed up creating custom layouts.

\subsection{Example Use Case (Continued)}\label{sec:example2}
We extend the example use case (Figure~\ref{fig:teaser}, \S\S\ref{sec:example0} and~\ref{sec:example1}) using selections and detailed views.
In the \emph{overlap} view in Figure~\ref{fig:teaser}B, we notice that in the rightmost column (neighborhood size 50), the bin for overlap of size seven contains a relatively high number of objects, based on its darker blue color.
These objects have a large difference between the embeddings (they share only 7 of their 50 nearest neighbors).
We select this bin (marked by the orange arrow in Figure~\ref{fig:teaser}B) by clicking on it, which adds all of its objects to the selection set.
This highlights those objects across all other views.

At first, we are interested in how the selected documents are distributed within both embeddings.
For this, we view the selection in the \emph{spread} view (Figure~\ref{fig:teaser}C).
It shows that up until neighbor 7, part of the selection has neighbors that stay close, while the neighbors of others move far away (the highlights are spread out across the vertical axis).
In addition, we can see neighbors from the high-dimensional embedding stay either close or move far away (left columns have highlights either on the bottom or the very top).
This means that some of them are not represented accurately in the 2D space.

The views in Figure~\ref{fig:big_picture} provide more detail on the selection.
Figure~\ref{fig:big_picture}a shows that for all of the selected instances, neighbors starting from neighbor 4 in the 2D embeddings are far away.
This means that most of the documents have few close neighbors (or none).
Figure~\ref{fig:big_picture}b shows that our selections are in low to medium density areas in both embeddings.
None of the selected examples have a particularly high local density.

One possible explanation for this is that documents that cannot be placed close to their original neighbors from the high-dimensional embedding end up as outliers in the lower dimensional one.

Next, we take a closer look at the selection.
The \emph{selection list} view (Figure~\ref{fig:teaser}D) shows that our selection set has 28 objects (on the bottom right).
After adding titles and publication year as meta data, we skim the documents, and find that they are from a wide range of topics.
In Figure~\ref{fig:teaser}E, we can see the diversity of our selection in terms of neighborhood retention.
Position encodes the ranks of the high dimensional embedding, and color the ones for the 2D embedding.
We can see that they differ widely with respect to neighbor correspondence within the 10 nearest neighbors.
While some of them keep many of their closest neighbors, others lose almost all of them (as can be seen by the orange-brown and pink colored boxes).
The latter set of documents have a less accurate representation of their immediate neighborhood in the low dimensional embedding.
We hover them, which causes them to highlight in other views, and explore some more details about them.

When viewing their position in the scatterplot view of the 2D embedding (Figure~\ref{fig:big_picture}c), we learn that they are positioned in the vicinity of clusters.
The other members of those clusters, however, are mostly not their original neighbors from the high dimensional embedding.
Viewing the neighborhoods in the \emph{neighborhood sequence} view, we can also see that some of them keep a few close neighbors.
Figure~\ref{fig:buddyplot} shows an example of this.
Moving some documents far away from their original neighbors is a result of tSNE that we do not want in our final 2D representation.
We can now test and compare other representations to see if they provide a better representation of those objects.

\section{Implementation}\label{sec:implementation}
We have implemented our approach in \sysname, a web-based prototype system for embedding comparison.
\sysname takes pairs of embeddings with the same (or overlapping) objects as input.
When the embeddings contain different objects, for example, word embeddings for different languages, \sysname requires a list of correspondences between the objects in both embeddings.

The system's frontend uses vue.js for the UI, and d3.js for the visualization components.
State management is implemented using the vuex addon to vue.js.
The backend is implemented in python3, using numpy with the Intel MKL library to speed up vector processing.
Analysis methods are implemented using the python sklearn library.

For efficient interaction, 100 nearest neighbors per embedding for each object are pre-computed using the scikit-learn nearest neighbor framework\footnote{https://scikit-learn.org/stable/modules/neighbors.html}, which automatically selects a tree data structure for fast neighborhood selection based on dimensionality and size of the dataset.
Memory usage of the tree is in $O(n)$, while looking up the neighbors of $n$ items will take $O(n\log n)$ time.
Once the neighborhoods are computed, we store this neighborhood transform as a matrix. 
It allows us to quickly compute the overview metrics that involve object neighborhoods, and save time for looking up neighborhood the next time the data set is loaded into \sysname.
While the number of neighbors is a system parameter that can be adapted, we found that 100 works well for all the embeddings we are working with in this work.
Since embeddings can be huge, all of the embedding data, including meta data for the objects is stored in the backend and transferred to the frontend and cached as needed.

\section{Use Cases}\label{sec:use_cases}
Previously, we have discussed an example for network node embeddings (\S\ref{sec:motivating_example}) and quality control of dimensionality reductions (\S\S\ref{sec:example0}, \ref{sec:example1}, \ref{sec:example2}).
This section introduces two additional use cases that illustrate the workflow of \sysname.
The first one compares two different document embeddings and the second one two different word embeddings.

\subsection{Document Embedding: Abstracts and Full Texts}
	\label{sec:abstract_vs_full}
	Before text collections can be used as input for machine learning algorithms or further analysis, they need to be embedded into a vector space.
The choice of the right algorithm and dataset is essential to preserve information relevant for the given task.

In this use case, we explore the differences created by two different embeddings.
This is a variation of the \exhypp scenario (\S\ref{sec:example_scenarios}).
The dataset is a corpus of 1477 publications from recent visualization conferences.
The first embedding was built using the abstracts as input and NMF~\cite{Sra2006} as embedding method.
The second embedding uses the full texts instead of abstracts.
We want to decide if it is worth the effort to obtain the text from pdfs instead of using the publicly available abstracts.
Using metrics based on local features to compare the two embeddings on a global level helps us understand the differences between the embeddings created from these two data sources.

\begin{figure*}[h!]
	\centering
	\includegraphics[width=.9\textwidth]{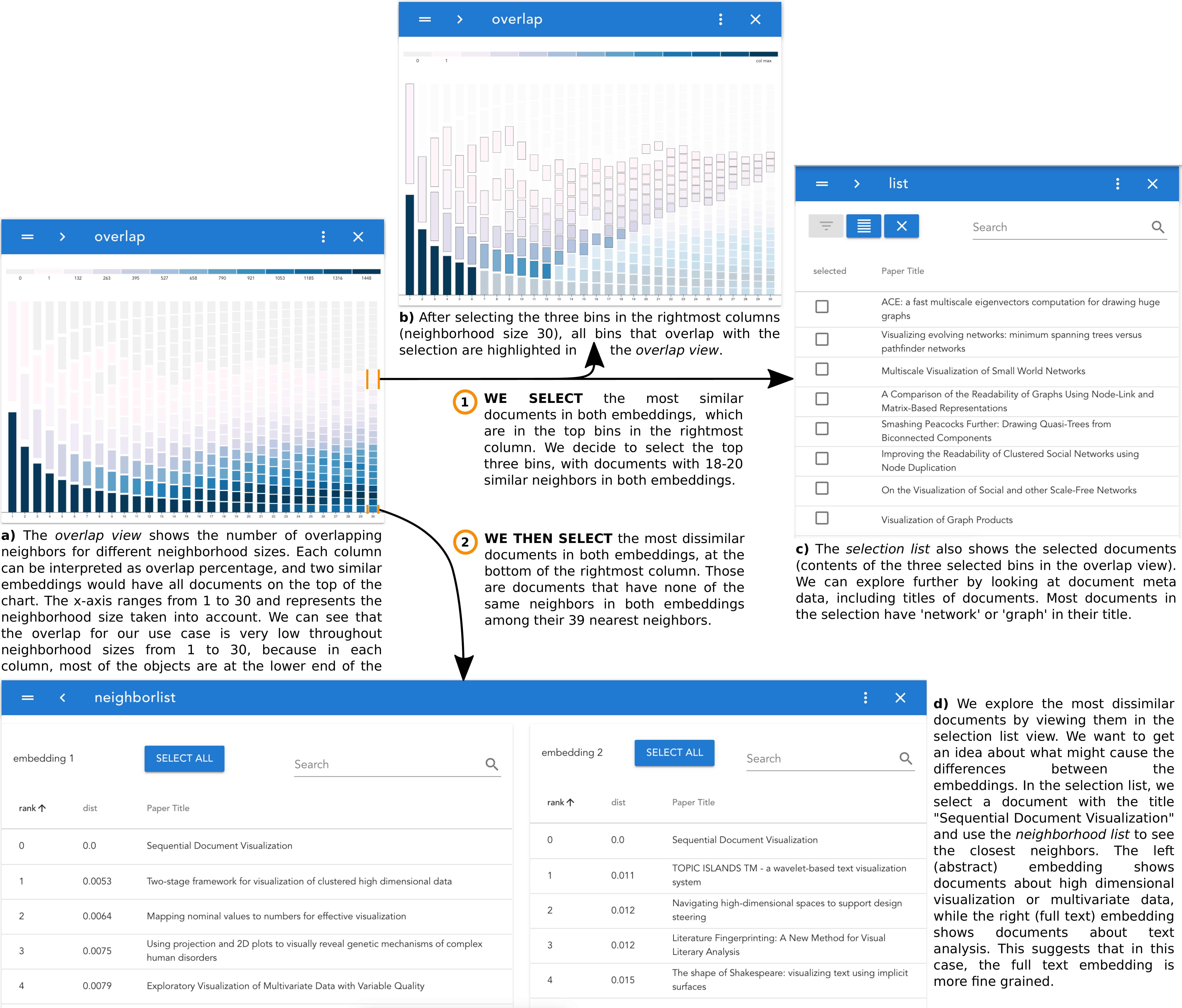}
	\caption{
		The views and sequence of selection for comparing two document embeddings generated on different source data.
	}
	\label{fig:uc1}
\end{figure*}

We start by assessing the global similarity between both models by looking at the local neighborhood overlap (Figure~\ref{fig:uc1}a).
The \emph{overlap view} allows us to see the overlap between both embeddings based on different neighborhood sizes.
We can see that most documents are on the lower end of the chart (dark blue color at the bottom) and have therefore very different neighbors.
Through mouse-over of the first column, tooltips show us that 29 items have the same first neighbor in both embeddings and 1448 items have a different first neighbor.

Even though most documents are placed in different neighborhoods, in each embedding, there are a few with similar neighbors in both.
We can select them using the \emph{overlap view} by clicking the top boxes in a larger neighborhood size column (Figure~\ref{fig:uc1}b).
In our example, the largest neighborhood size is 30 and visible in the rightmost column.
The selected three boxes include documents with 20 (the top box), 19 (middle box), and 18 (lowest box) similar neighbors out of 30 possible.
The \emph{selection list} now allows us to look at the selected documents and read the titles of the publications.
This reveals that documents with the most similar neighborhoods talk about networks or graphs (Figure~\ref{fig:uc1}c).

We saw at the beginning that most documents have very different neighborhoods.
We start exploring them by selecting the documents with the lowest overlap given a neighborhood size of 30.
All selected publications have no overlap within the first 30 neighbors.
We can compare the neighborhood of different publications by using the \emph{neighborhood list}.
By looking at the publication with the title "Sequential Document Visualization", we can see that the closest neighbors in the abstract embedding are high dimensional visualization and multivariate data publications.
The closest neighbors in the full text embedding are about text visualization (Figure~\ref{fig:uc1}d).
Text data is a specific case of high dimensional data, which suggests that the full text embedding has a higher resolution and can represent more specific topics.
Another example is the publication titled ``Interactive Multiscale Tensor Reconstruction for Multiresolution Volume Visualization''.
The closest neighbors in the abstract embedding are volume rendering publications, while the full text neighbors are about tensors.
Again, this pattern revealed through exploration suggests that the full text embedding represents topics at a higher resolution compared to the abstract embedding.

To understand the differences between the embeddings on a more global scale, we use the \emph{scatterplot} view.
UMAP is used to reduce the dimensionality of both embeddings, and we start looking at the clusters in both embeddings by using the rectangular selection tool with the mouse to select and inspect different clusters in both embeddings.
This shows us how documents that are close in one embedding disperse across the second embedding.
We can see that many clusters selected in the abstract embedding are represented in the full text embedding as multiple small clusters that stay close to each other.
This suggests that the full text embedding groups similar documents as the abstract embedding, but with a higher resolution.
\subsection{Word Embedding: Changes in Language}
	\label{sec:eebo}
	We are collaborating with literary linguists who are interested in how the English language has changed over the centuries.
To explore these changes, we have constructed word vector embeddings from two different corpora: a historic corpus consisting of over 50,000 books published between 1475 to 1700 from the Early English Books Online (EEBO) collection as transcribed by the Text Creation Partnership (TCP)~\cite{eebo-tcp} and a modern corpus built from the English language Wikipedia.
This is a variation of scenario \excorpora (\S\ref{sec:example_scenarios}).
We expect that changes in word meanings between the two eras would be reflected in the differences between their respective word vector embeddings.
For the experiments described here, we built the embeddings using GloVe with 300 dimensions, a window size of 15, and a minimum token count of 5.

\begin{figure*}
	\centering
	\includegraphics[width=\textwidth]{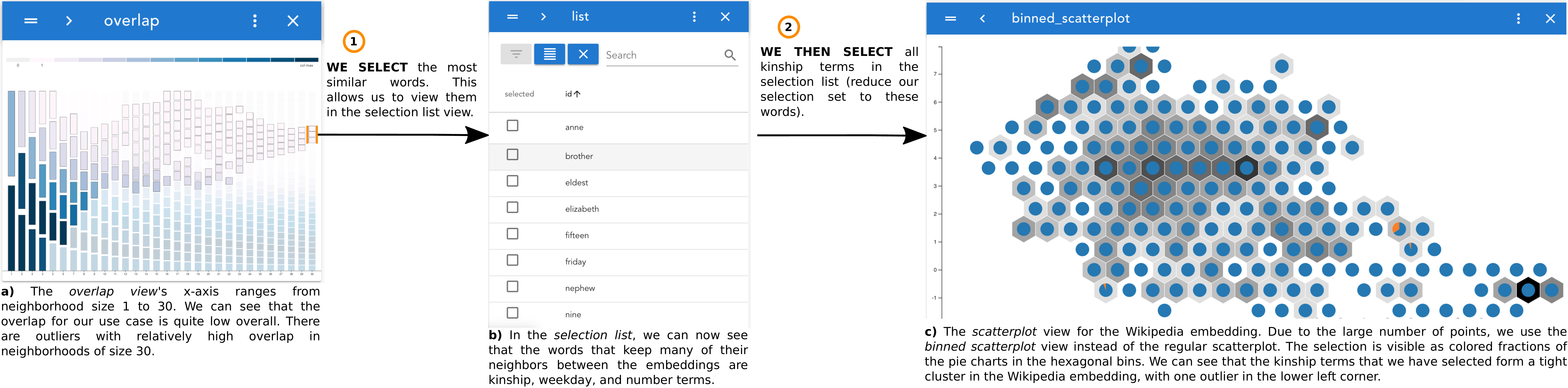}
	\caption{
		The views and sequence of selections for comparing two word embeddings generated on two different datasets.
	}
	\label{fig:uc2}
\end{figure*}

As word vector embeddings model words based on their context in a large corpus, words that are used in similar context are closer in the resulting space than words that have different usage patterns.
We can use this to study changes in word meaning based on the local neighborhood of words.
We use the \emph{overlap} view (Figure~\ref{fig:uc2}a) to look at the global similarity based on the local neighborhoods up to 30 neighbors.
By looking at the color fields in each column, we can see that most of the words are in bins in the lower area of the chart.
This means that the overall neighborhood overlap is low.
Most words have different neighbors.
In contrast, a few words seem to similar neighborhoods (on the top of the view).
Those catch our attention, and we want to learn more about them.
By selecting the top boxes in the most right column, we select the words with the highest overlap between both embeddings (Figure~\ref{fig:uc2}a).
We use the \emph{selection list} view to look at the selected words (Figure~\ref{fig:uc2}b), and can see that the most similar words are weekdays, number words, names or kinship relations.

The linguists were not surprised by that finding as these words have not changed in usage or meaning. Seeing their knowledge reflected in the embeddings and our tool increased their confidence in both.

We take a closer look at these semantically close set of words, and confirm by connecting back to the global context of the embeddings through the \emph{neighborhood distance} view and the \emph{density} view that those are clusters that are particularly close within both embeddings.

An interesting group of stable words are kinship terms, which we select using the selection list view.
We use the \emph{binned scatterplot} view (Figure~\ref{fig:uc2}c) to see the position of those terms in both embeddings.
We can see that the terms form tight clusters in both embeddings.

We can see one outlier in the Wikipedia embedding, the term \emph{parent.}
Interested in how the usage of the term changed, we look at the local neighbors of the term in each embedding using the \emph{neighbor list} view.
The closest neighbors in the Wikipedia embedding are subsidiary, ownership, and acquisition.
This reflects the meaning of parent company.
In contrast, the closest neighbors in the TCP embedding are child, mother, and father.
This reflects less of a word usage change and more of a bias in the Wikipedia dataset.

Another way to find more interesting word usage changes is by selecting the words with the lowest overlap between both embeddings using the \emph{overlap} view.
Our collaborators used this fast way to validate ideas they had already in mind and to find unknown changes.
Some examples are: the word \emph{cup} changes the usage from a drinking vessel to a trophy.
The word \emph{crisis} was used to describe illness and disease in TCP and changed to financial and conflict in general in modern English.
The word \emph{forum} changed from market, theater, and senate to discussion, global and conferences.

After the initial sessions, we had additional meetings with the experts.
They were very interested in comparing more historic and contemporary corpora, including subsets of TCP and embeddings built from the Google ngrams corpus.
During those comparisons, they were able to uncover a range of additional insights about change in the English language over time.
One striking example for our collaborators was the change of the term \emph{conduct} from "to lead something" to "behavior".

\section{Discussion}\label{sec:discussion}
In this paper, we have presented an approach to comparing object embeddings.
We surveyed a range of potential scenarios to abstract common challenges (\S\ref{sec:example_scenarios}) and identify solution strategies (\S\ref{sec:strategies}).
To help address these challenges, we designed \sysname (\S\ref{sec:system_design}), that implements the strategies to help users with comparing pairs of embedding models.

\subsection{Expert Feedback}
We discussed the system and the design of the views with three expert users.
All of them are language experts and were interested in or had experience working with word vector embeddings.
Two of them are interested in language change within the same language over time.
For this, they are using word embeddings to identifying subsets of words whose meaning differs in interesting ways between different historic versions of a language.
The third expert, a cognitive scientist, is interested in how associations between words and their meaning differ between different languages.
He is using word embeddings for different languages with translation links to study those differences.

During each of the three sessions we conducted with each of the experts, we were in control of the system, while the analysis was guided by the expert.
We started each of the sessions by explaining the metrics and the visual encoding of the overviews.
Initially, we showed simpler configurations with single distributions and explained the visual encodings with practical examples from embeddings that the experts were familiar with.
We then extended the overview displays to gradually include more information, explaining and motivating the visual encodings with the example.
During this training, experts could ask questions and discuss interpretations of the visualizations and the underlying embeddings.
With us in direct control of \sysname, the experts were able to effectively interpret and use the visualizations, and direct us towards interesting examples and findings in their data.
Many of them were received enthusiastically by the experts.
We provided screen shots of interesting configurations, and the experts used those to discuss some of the findings with their colleagues.

In addition to the findings that experts were able to uncover with \sysname, we also received direct feedback about the usability of our system and the interpretability of the visual encodings.
While they generally found the visualizations straightforward to work with, two of the experts remarked that the overview visualizations take some time initially to learn to read and interpret.
This was, in particular, mentioned for the overlap view.
They expressed concern that language scholars in general might have a hard time using the displays effectively without explicit guidance.

Two of the experts further suggested improving the presentation of metadata attributes in tables and tooltips.
This could be done by adding visual summaries for a number of columns, and providing context for single values in form of distributions for the entire dataset.
Beyond those additions to help work more effectively with metadata, the third expert (cognitive scientist) was interested in showing summaries of metadata values for each bin in the overview visualizations.
This would allow him to correlate additional linguistic metrics about the words in the embeddings.

This feedback suggests that users are able to interpret the visual metaphors in conjunction with the rest of the \sysname after some initial training and with adequate guidance.
To improve the usability of the system for users interested in comparing embeddings, we added some initial guidance that helps with setting up relevant views for the scenario at hand (\S\ref{sec:view_selection}).
In addition, we are planning to develop and implement more advanced guidance mechanisms for complex multi-view systems.
They will be integrated into \sysname to help guide users through meaningful analyses of their embeddings, without the help of a \sysname expert.

\subsection{Limitations}
While our prototype system shows promise in its utility for comparing embeddings, it also emphasizes some limitations in the current state of our work.
Foremost, we have only considered pairwise comparison.
Multi-way comparisons are common, and are particularly important for comparing different hyper-parameter settings or to determine the consistency of embedding algorithms across multiple runs.
Unfortunately, extensions of our methods beyond pairwise comparison are challenging: many of the metrics and visual designs do not scale beyond pairs.

Another limitation of our approach is its complexity: the approach relies on a collection of different metrics and views.
Users must be able to understand the different options and how to use them together to achieve their goals.
We believe this issue may be addressed by providing more structured interactions and guidance to help users apply views that address their goals.
Using our implementation to work with experts on a diverse set of scenarios helps us gather feedback to improve interaction design and user guidance.

Our approach has focused on top down analysis: overviews are used to assess overall differences and identify selections to be explored in more detail.
Bottom up analyses do occur.
For example, the literature scholars like to begin by asking about the behaviors of specific words, which they then use to generalize to broader sets.
Such workflows are possible with our current system, but may be better supported with more tailored views and interactions. More generally, the task list we considered is not exhaustive; supporting more tasks may require adding new views and interactions.

Our approach is domain agnostic, treating embeddings as abstract collections of distances.
While this allows our tool to be broadly applicable, it also misses opportunities to use properties of the domain to enhance the embeddings.
For example, knowing that the objects are documents or images suggests different operations to perform with them - at a minimum, they can be displayed.
A related limitation is that our approach makes limited use of metadata about the objects.
It is possible to select semantic groups using the selection list, for example, to select documents labeled with a particular genre or words labeled with a particular part of speech.
However, future work could facilitate comparisons between known groupings of objects, for example, to determine if different genres behave similarly in different embeddings.

Our summary views aggregate data and can scale to large collections of objects.
We use subset selection and scanning to reduce the problem size for detailed analysis.
In practice, our system works with collections of several thousand objects - for the examples in this paper, we have reduced the word embeddings to the 20,000 most common words.
Scaling to larger object collections, which are not uncommon in machine learning applications, will be challenging for performance reasons, but also in providing mechanisms to help refine large selections when boxes in the overviews may represent hundreds or thousands of objects.

\section*{Acknowledgments}
This work was funded in part by NSF Award 1841349 and DARPA award  FA8750-17-2-0107.
We thank our domain collaborators including Michael Witmore, Jonathan Hope, and Gary Lupyan.

\bibliographystyle{abbrv-doi}

\bibliography{main}

\begin{IEEEbiography}
	[{\includegraphics[width=1in,height=1.25in,clip,keepaspectratio]{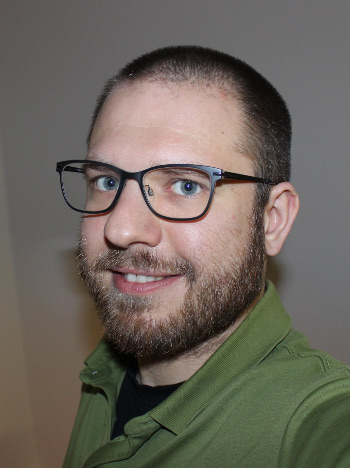}}]{Florian Heimerl}
	is a postdoctoral researcher in the visualization group at the University of Wisconsin--Madison. He received his PhD in computer science and his Diplom degree in computational linguistics from the University of Stuttgart, Germany. In Stuttgart, he was a doctoral student at the Institute for Visualization and Interactive Systems. His research interests include information visualization, visual analytics, and visual text analysis.
\end{IEEEbiography}

\begin{IEEEbiography}
	[{\includegraphics[width=1in,height=1.25in,clip,keepaspectratio]{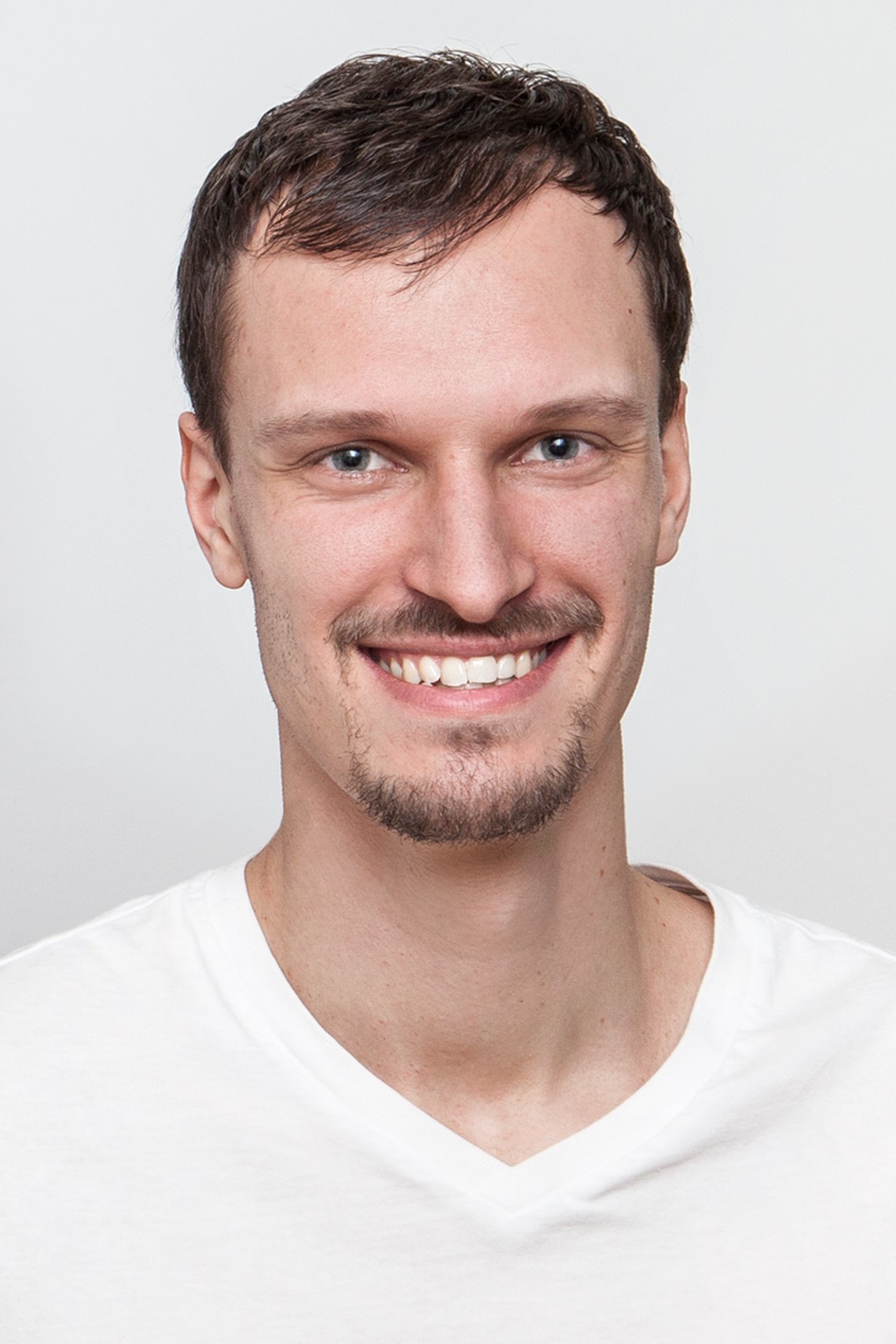}}]{Christoph Kralj}
	is a Ph.D. student of computer science at the University of Vienna, Austria, since 2017. He received his MSc in Computational Sciences from the University of Vienna in 2016 and a BSc in Earth Sciences from Technical University of Graz, Austria.
	His research interest is about bridging the gap between human understanding and computational solutions with a strong focus on textual data. He used tools and methodologies from natural language processing, visualization, machine learning, and human-computer interaction to create interactive systems.
\end{IEEEbiography}

\begin{IEEEbiography}
	[{\includegraphics[width=1in,height=1.25in,clip,keepaspectratio]{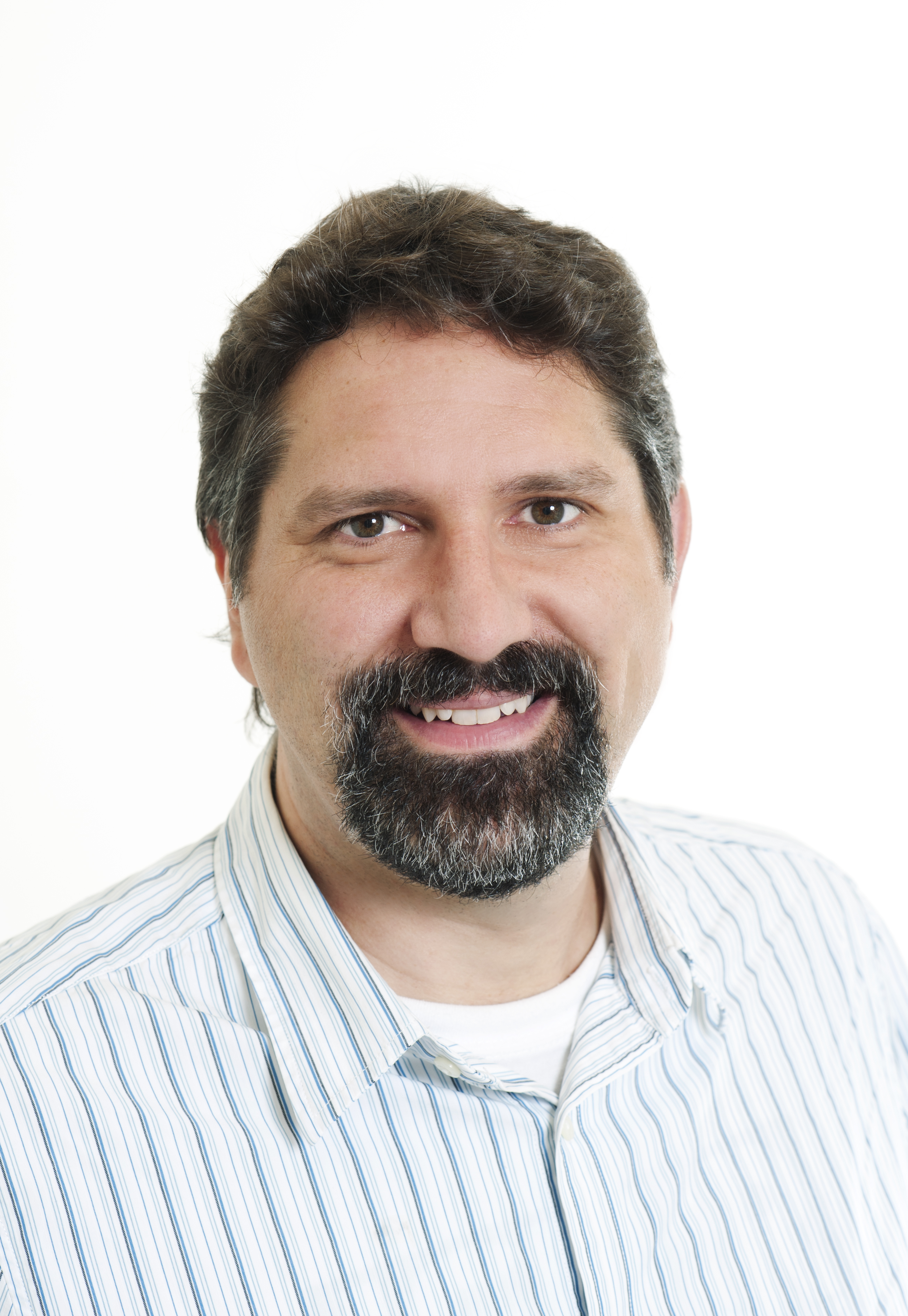}}]{Torsten M\"oller}
	is a professor of computer science at the University of
	Vienna, Austria, since 2013. Between 1999 and 2012 he served as a
	Computing Science faculty member at Simon Fraser University, Canada. He
	received his PhD in Computer and Information Science from Ohio State
	University in 1999 and a Vordiplom (BSc) in mathematical computer science
	from Humboldt University of Berlin, Germany. He is a senior member of IEEE
	and ACM, and a member of Eurographics. His research interests include
	algorithms and tools for analyzing and displaying data with principles
	rooted in computer graphics, human-computer interaction, signal
	processing, data science, and visualization.
\end{IEEEbiography}

\begin{IEEEbiography}[{\includegraphics[width=1in,clip,keepaspectratio]{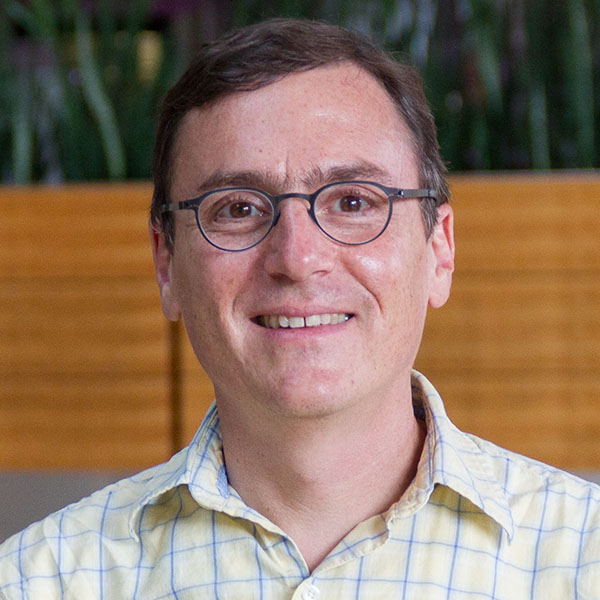}}]{Michael Gleicher}
	is a Professor in the Department of Computer Sciences at the University of Wisconsin, Madison.  Prof. Gleicher is founder of the Department's Visual Computing Group. He co-directs both the Visual Computing Laboratory and the Collaborative Robotics Laboratory at UW-Madison. His research interests span the range of visual computing, including data visualization, robotics, and virtual reality. Prior to joining the University, Prof. Gleicher was a researcher at The Autodesk Vision Technology Center and in Apple Computer's Advanced Technology Group. He earned his Ph. D. in Computer Science (1994) from Carnegie Mellon University, and holds a B.S.E. in Electrical Engineering from Duke University (1988). In 2013-2014, he was a visiting researcher at INRIA Rhone-Alpes. Prof. Gleicher is an ACM Distinguished Scientist.
\end{IEEEbiography}

\end{document}